\documentclass[prd,amsmath,showpacs,preprintnumbers]{revtex4}
\usepackage{epsfig,dcolumn}

\newcommand{\beq}{\begin{equation}}
\newcommand{\eeq}{\end{equation}}

\newcommand{\rf}[4]{#1 {\bf #2}, #3 (#4)}

\newcommand{\pr}{Phys.\ Rev.}

\newcommand{\physl}{Phys.\ Lett.}

\newcommand{\np}{Nucl.\ Phys.}
\newcommand{\npps}[3]{Nucl.\ Phys.\ {\bf B} (Proc.\ Suppl.) #1, 
	#2 (#3)}

\newcommand{\etal}{\emph{et al.}}

\newcommand{\del}{\partial}
\newcommand{\adj}{\dagger}

\newcommand{\qhat}{\ensuremath{\hat{q}}}
\newcommand{\muhat}{\hat{\mu}}

\newcommand{\lapI}{\ensuremath{\partial^2\text{(I)}}}
\newcommand{\lapII}{\ensuremath{\partial^2\text{(II)}}}
\newcommand{\lapIII}{\ensuremath{\partial^2\text{(III)}}}

\newcommand{\oa}[1]{\ensuremath{{\cal O}(a^{#1})}}

\begin{document}

\preprint{FSU-CSIT-02-17}
\preprint{ADP-02-75/T514}

\title{Gluon Propagator on Coarse Lattices in Laplacian Gauges}

\author{Patrick O.~Bowman}
\author{Urs M.~Heller}
\affiliation{Department of Physics and 
School for Computational Science and Information Technology, 
Florida State University, Tallahassee FL 32306-4120, USA}
\author{Derek B.~Leinweber}
\author{Anthony G.~Williams}
\affiliation{Special Research Centre for the Subatomic Structure of Matter 
and \\ The Department of Physics and Mathematical Physics,
University of Adelaide, Adelaide, SA 5005, Australia}

\begin{abstract}
The Laplacian gauge is a nonperturbative gauge fixing that reduces to 
Landau gauge in the asymptotic limit.  Like Landau gauge, it respects
Lorentz invariance, but it is free of Gribov copies; the gauge 
fixing is unambiguous.  In this paper we study the infrared behavior of
the lattice gluon propagator in Laplacian gauge by using a variety of 
lattices with spacings 
from $a = 0.125$ to 0.35 fm, to explore finite volume and discretization
effects.  Three different implementations of the Laplacian gauge are defined 
and compared.  The Laplacian gauge propagator has already been claimed to be 
insensitive to finite volume effects and this is tested on lattices with large
volumes. 
%
\end{abstract}

\pacs{ 
12.38.Gc  
11.15.Ha  
12.38.Aw  
14.70.Dj  
}

\maketitle

\section{Introduction}

The lattice provides a useful tool for studying the gluon propagator because
it is a first principles treatment that can, in principle, access any momentum
window.  
There is tremendous interest in the infrared behavior of the 
gluon propagator as a probe into the mechanism of confinement~\cite{background}
and lattice studies focusing on its ultraviolet behavior have been used to 
calculate the running QCD coupling~\cite{Bec99}.  Such studies can also inform 
model hadron calculations~\cite{models}.  Although there has 
recently been interest in Coulomb gauge~\cite{Cuc01} and generic covariant
gauges~\cite{Giu01}, the usual gauge 
for these studies has been Landau gauge, because it is a (lattice) Lorentz
covariant gauge that is easy to implement on the lattice, and its popularity 
means that results from the lattice can be easily compared to studies that use 
different methods.  Finite volume effects
and discretization errors have been extensively explored in lattice Landau
gauge~\cite{Lei99,Bon00,Bon01}.  
Unfortunately, lattice Landau gauge suffers from the well-known problem of 
Gribov copies.  Although the ambiguity originally noticed by 
Gribov~\cite{Gri78} is not present on the lattice, the
maximization procedure used for gauge fixing does not uniquely fix the gauge.
There are, in general, many local maxima for the algorithm to choose from, 
each one corresponding to a Gribov copy, and no local algorithm can choose the
global maximum from among them.  While various remedies have
been proposed~\cite{Het98,evolutionary}, they are either unsatisfactory or 
computationally very intensive.  For a recent discussion of the Gribov problem
in lattice gauge theory, see Ref.~\cite{Wil02}.

An alternative approach is to operate in the so-called 
Laplacian gauge~\cite{Vin92}.  This gauge is ``Landau like'' in that it has
similar smoothness and Lorentz invariance properties~\cite{Vin95}, but it 
involves a non-local gauge fixing procedure that avoids lattice Gribov copies.
Laplacian gauge fixing also has the virtue of being rather faster than
Landau gauge fixing on the lattice.
The gluon propagator has already been studied in Laplacian gauge in 
Refs.~\cite{Ale01,Ale02} and the improved staggered quark
propagator in Laplacian gauge in Ref.~\cite{quarkprop}.

In this report we explore three implementations of the Laplacian gauge and
their application to the gluon
propagator on coarse, large lattices, using an improved action as has 
been done for Landau gauge in Ref.~\cite{Bon01}. 
We study the gluon propagator in quenched QCD (pure $SU(3)$ 
Yang-Mills), using an \oa{2} mean-field improved gauge action.
To assess the effects of finite lattice spacing, we calculate the
propagator on a set of lattices from $10^3\times 20$ at $\beta = 3.92$
having $a = 0.353$ fm to $16^3 \times 32$ at $\beta = 4.60$ having 
$a = 0.125$ fm.  To assist us in observing possible finite volume
effects, we add to this set a $16^3 \times 32$ lattice at $\beta =
3.92$ with $a = 0.353$, which has the very large physical size of
$5.65^3 \times 11.30 \text{ fm}^4$.

The infrared behavior of the Laplacian gauge gluon propagator is found to be
qualitatively similar to that seen in Landau gauge.  Like 
Refs.~\cite{Ale01,Ale02}, little volume dependence is seen in the propagator, 
but, unlike Landau gauge, the Laplacian gauge displays strong sensitivity to 
lattice spacing, making large volume simulations difficult.
We conclude that further work involving an improved gauge fixing is desired.

\section{The Laplacian Gauges}

Laplacian gauge is a nonlinear gauge fixing that respects rotational
invariance, has been seen to be smooth, yet is free of Gribov ambiguity.
It reduces to Landau gauge in the asymptotic limit, yet is computationally 
cheaper than Landau gauge.  There is, however, more
than one way of obtaining such a gauge fixing in $SU(N)$. 
The three implementations of Laplacian gauge fixing discussed are
\begin{enumerate}
\item \lapI\ gauge (QR decomposition), used by Alexandrou \emph{et al.\/} in 
	Ref.~\cite{Ale01}.
\item \lapII\ gauge (Maximum trace), where the Laplacian gauge
        transformation is projected  
	onto $SU(3)$ by maximizing its trace.  Also used in 
	Ref.~\cite{quarkprop}.
\item \lapIII\ gauge (Polar decomposition), the original prescription described
 	in Ref.~\cite{Vin92} and tested in Ref.~\cite{Vin95}.
\end{enumerate}
All three versions reduce to the same gauge in $SU(2)$.

The gauge transformations employed in Laplacian gauge fixing are constructed 
from the lowest eigenvectors of the covariant lattice Laplacian
\begin{equation}
\sum_y \sum_j \Delta(U)(x,y)^{ij} v(y)_j^s = \lambda^s v(x)_i^s,
\end{equation}
where
\begin{equation}
\Delta(U)(x,y)^{ij} \equiv \sum_\mu \bigl[ 2\delta(x-y) \delta^{ij}
   - U_\mu(x)^{ij} \delta(x+\muhat-y) - U_\mu(y)^{\adj ij} 
   \delta(y+\muhat-x) \bigr],
\end{equation}
$i,j = 1,\ldots,N$ for $SU(N)$ and $s$ labels the eigenvalues and eigenvectors.
Under gauge transformations of the gauge field,
\begin{equation}
U_\mu(x) \rightarrow U_\mu^G(x) = G(x) U_\mu(x) G^\adj(x+\mu),
\end{equation}
the eigenvectors of the covariant Laplacian transform as 
\begin{equation}
v(x)^s \rightarrow G(x) v(x)^s
\end{equation}
and this property enables us to construct a gauge fixing that is independent
of our starting place in the orbit of gauge equivalent configurations.

The three implementations discussed differ in the way that the gauge 
transformation is constructed from the above eigenvectors.  In all cases
the resulting gauge should be unambiguous so long as the Nth and (N+1)th 
eigenvectors are not degenerate and the eigenvectors can be satisfactorily
projected onto $SU(N)$.  A complex $2\times 2$ matrix can be uniquely projected
onto $SU(2)$, but this is not the case for $SU(N)$.  Here we can think of 
the projection method as defining its own, unambiguous, Laplacian gauge fixing.

In the original formulation~\cite{Vin92}, which we shall rather perversely 
refer to as \lapIII, the lowest $N$ eigenvectors are required to gauge fix
an $SU(N)$ gauge theory.  These form the columns of a complex $N\times
N$ matrix, 
\begin{equation}
M(x)^{ij} \equiv v(x)_i^j
\end{equation}
which is then projected onto $SU(N)$ by polar decomposition.  Specifically, it
is possible to express $M$ in terms of a unitary matrix and a positive 
hermitian matrix: $M = UP$.  This decomposition is unique if 
$P = (M^\adj M)^{1/2}$ is invertible, which will be true if $M$ is 
non-singular, i.e., if the eigenvectors used to construct
$M$ are linearly independent.
The gauge transformation $G(x)$ is then
obtained by factoring out the determinant of the unitary matrix
\begin{equation}
G^\adj(x) = U(x) / \det[U(x)].
\end{equation}
The gauge transformation $G(x)$ obtained in this way is used to transform the 
gauge field (i.e., the links) to give the Laplacian gauge-fixed gauge field.  
$G(x)$ can be uniquely defined by this presciption except on a set of gauge 
orbits with measure zero (with respect to the the gauge-field functional 
intregral).  Note that if we perform a random gauge transformation $G_r(x)$ on 
the initial gauge field used to define our Laplacian operator, then we will 
have $v(x)^s \to v'(x)^s = G_r(x) v(x)^s$ and $M(x) \to M'(x) = G_r(x) M(x)$.  
We see that $P \equiv (M^\dagger M)^{1/2} \to P'=P$ and hence 
$G(x) \to G'(x)=G(x)G^\dagger_r(x)$.  When this is applied to the transformed 
gauge field it will be taken to exactly the same point on the gauge orbit as 
the original gauge field went to when gauge fixed.  Thus all points on the 
gauge orbit will be mapped to the same single point on the gauge orbit after 
Laplacian gauge fixing and so it is a complete (i.e., Gribov-copy free)
gauge fixing.
This method was investigated for $U(1)$ and $SU(2)$~\cite{Vin95}.
It is clear that any prescription for projecting $M$ onto some $G^\dagger(x)$,
which preserves the property $G(x) \to G'(x)=G(x)G^\dagger_r(x)$ 
under an arbitrary gauge transformation $G_r(x)$, will also be a Gribov-copy
free Laplacian gauge fixing.  Every different projection method with this
property is an equally valid but distinct form of Laplacian gauge fixing.

The next approach was used in Ref.~\cite{Ale01}, and we shall refer to 
it as \lapI\ gauge.  There it was noted that only $N-1$ eigenvectors are 
actually required.  To be concrete, we discuss $N=3$.  First, apply a gauge 
transformation, $G(x)^1$, to the first eigenvector such that
\begin{equation}
\bigl[ G(x)^1 v(x)^1 \bigr]_1 = || v(x)^1 ||
\end{equation}
and
\begin{equation}
\bigl[ G(x)^1 v(x)^1 \bigr]_2 = \bigl[ G(x)^1 v(x)^1 \bigr]_3 = 0,
\end{equation}
where subscripts label the vector elements, i.e., the eigenvector - with 
dimension 3 - is rotated so that its magnitude is entirely in its first 
element.
Another gauge transformation, $G(x)^2$, rotates the second eigenvector, 
$v(x)^2$, such that
\begin{equation}
\bigl[ G(x)^2 v(x)^2 \bigr]_2 = \sqrt{(v_2^2)^2 + (v_3^2)^2},
\end{equation}
and
\begin{equation}
\bigl[ G(x)^2 v(x)^2 \bigr]_3 = 0.
\end{equation}
This second rotation is an $SU(2)$ subgroup, which does not act on $v_1^2(x)$.
The gauge fixing transformation is then $G(x) = G(x)^2 G(x)^1$.  Compare this
to QR decomposition, where a matrix, $A$, is rewritten as the product of an
orthogonal matrix and an upper triangular matrix.  The gauge transformations 
are thus like Householder transformations.

In addition, we explore a third version, \lapII gauge, where $G(x)$ is 
obtained by projecting $M(x)$ onto $SU(N)$ by trace maximization.  $M(x)$ is 
again composed of the $N$ lowest eigenvectors and the trace of $G(x)M(x)^\adj$ 
is maximized by iteration over Cabbibo-Marinari subgroups.

\section{The Gluon Propagator in Laplacian Gauge}

We extract the gluon field from the lattice links by
\begin{equation}
\label{eq:gludef}
A_\mu(x+\muhat/2) = \frac{1}{2ig_0u_0} \bigl\{ U_\mu(x) - U_\mu^\adj(x) 
	\bigr\}_{\text{traceless}},
\end{equation}
which differs from the continuum field by terms of \oa{2}.  $A_\mu$ is then
transformed into momentum space,
\begin{equation}
A_\mu(\qhat) = \frac{1}{V} \sum_x e^{-i\qhat \cdot (x+\muhat/2)}
	A_\mu(x+\muhat/2),
\end{equation}
where the available momenta, $\qhat$, are given by
\begin{equation}
\qhat_\mu  = \frac{2 \pi n_\mu}{a L_\mu}, \qquad
n_\mu \in  \Bigl( -\frac{L_\mu}{2}, \frac{L_\mu}{2} \Bigr].
\label{eq:qhat}
\end{equation}
$L_\mu$ is the number of lattice sites in the $\mu$ direction.  The momentum
space gluon propagator is then
\begin{equation}
D_{\mu\nu}^{ab}(\qhat) = \langle A_\mu^a(-\qhat) A_\nu^b(\qhat) \rangle.
\end{equation}
Note that this definition includes a factor of $u_0^{-2}$ from 
Eq.~\eqref{eq:gludef}; this is the same normalization that was used in 
Ref.~\cite{Bon01}.

In the continuum, the gluon propagator has the tensor structure
\footnote{Note that we have absorbed a factor of $q^{-2}$ into $F(q^2)$
compared to Ref.~\cite{Ale01}.}
\begin{equation}
D_{\mu\nu}^{ab}(q) = \bigl( \delta_{\mu\nu}-\frac{q_{\mu}q_{\nu}}{q^2} 
   \bigl) \delta^{ab}D(q^2) + \frac{q_{\mu}q_{\nu}}{q^2} \delta^{ab} F(q^2). 
\label{eq:tensor_struct}
\end{equation}
In Landau gauge the longitudinal part will be zero for all $q^2$, but this 
will not be the case in Laplacian gauge.
We note that 
\begin{equation}
\frac{q_\mu D_{\mu\nu}^{ab}(q) q_\nu}{q^2} = \delta^{ab}F(q^2)
\end{equation}
and use this to project out the longitudinal part.  This, unfortunately, makes
it impossible to directly measure the scalar propagator at zero four-momentum,
$D(0)$.
We are, however, able to measure the full propagator, $\mathcal{D}(0)$.  In
a covariant gauge $F(q^2) \rightarrow \frac{\xi_0}{q^2}$ where
$\xi_0$ is the bare gauge fixing parameter.

On the lattice the bare propagator is measured, which is related to the 
renormalized propagator by,
\begin{equation}
\label{eq:renorm}
D(q^2) = Z_3(\mu) \widetilde{D}(q^2;\mu^2) \qquad
F(q^2) = Z_3(\mu) \widetilde{F}(q^2;\mu^2) 
\end{equation}
where $\mu$ is the renormalization point.  In a renormalizable theory such as
QCD, the renormalized quantities become independent of the regularization
parameter in the limit where it is removed. 
$Z_3$ is then defined by some renormalization prescription, such as the
off-shell subtraction (MOM) scheme, where the renormalized propagator is 
required to satisfy,
\begin{equation}
\widetilde{D}(\mu^2) = \frac{1}{\mu^2}.
\end{equation}
It follows that 
\begin{equation}
q^2 D(q)|_{q^2=\mu^2} = Z_3(\mu;a).
\end{equation}
With covariant gauge fixing, the longitudinal part is usually treated by 
absorbing the renormalization into the gauge parameter, $\xi_0 = Z_3 \xi$.
We shall not discuss the renormalized propagator in this paper, but only
consider relative normalizations for the purpose of comparing different data
sets.

\begin{table}[t]
\begin{ruledtabular}
\begin{tabular}{lcccdr}
   & Dimensions	     & $\beta$ & $a\text{ (fm)}$ & 
\multicolumn{1}{c}{Volume $\text{(fm}^4\text{)}$} &
Configurations \\
\hline
1  & $12^3\times 24$ &   4.60  &   0.125  & 1.50^3 \times 3.00  & 100 \\
2i & $16^3\times 32$ &   4.60  &   0.125  & 2.00^3 \times 4.00  & 100 \\
2w & $16^3\times 32$ &   5.85  &   0.130  & 2.08^3 \times 4.16  &  80 \\
3  & $16^3\times 32$ &   4.38  &   0.166  & 2.64^3 \times 5.28  & 100 \\
4  & $12^3\times 24$ &   4.10  &   0.270  & 3.24^3 \times 6.48  & 100 \\
5  & $10^3\times 20$ &   3.92  &   0.353  & 3.53^3 \times 7.06  & 100 \\
6  & $16^3\times 32$ &   3.92  &   0.353  & 5.65^3 \times 11.30 &  89 \\
\end{tabular}
\end{ruledtabular}
\caption{\label{tab:latlist}Details of the lattices used to calculate the 
gluon propagator.  
Lattice 2w was generated with the Wilson gauge action.}
\end{table}

The ensembles studied are listed in Table~1. 
To help us explore lattice artifacts, some of the following figures will 
distinguish data on the basis of its momentum components.
Data points that come from momenta lying entirely along a spatial
Cartesian direction are indicated with a square while points from
momenta entirely in the temporal direction are marked with a triangle.  As
the time direction is longer than the spatial directions any difference
between squares and triangles may indicate that the propagator is
affected by the finite volume of the lattice.  Data points from
momenta entirely on the four-diagonal are marked with a diamond.
A systematic separation of data points taken on the diagonal from those
in other directions indicates a violation of rotational symmetry.

In the continuum, the scalar function is rotationally
invariant.  Although the hypercubic lattice breaks O(4) invariance, it does 
preserve the subgroup of discrete rotations Z(4).  In our case, this symmetry
is reduced to Z(3) as one dimension will be twice as long as the other three
in each of the cases studied.  We exploit this discrete rotational symmetry to 
improve statistics through Z(3) averaging~\cite{Lei99,Bon01}. 

As has become standard practice in lattice gluon propagator studies, we 
select our lattice momentum in accordance with the tree-level behavior of
the action.  With this improved action,
\begin{equation}
q_\mu = \frac{2}{a}\sqrt{ \sin^2 
	\Bigl( \frac{\qhat_\mu a}{2} \Bigr)
	+ \frac{1}{3}\sin^4 \Bigl( \frac{\qhat_\mu a}{2} \Bigr) 
	};
\end{equation}
this is discussed in more detail in Ref.~\cite{Bon01}.

\section{Results}
\subsection{Finer lattices}

\begin{figure}[h]
\begin{center}
\epsfig{figure=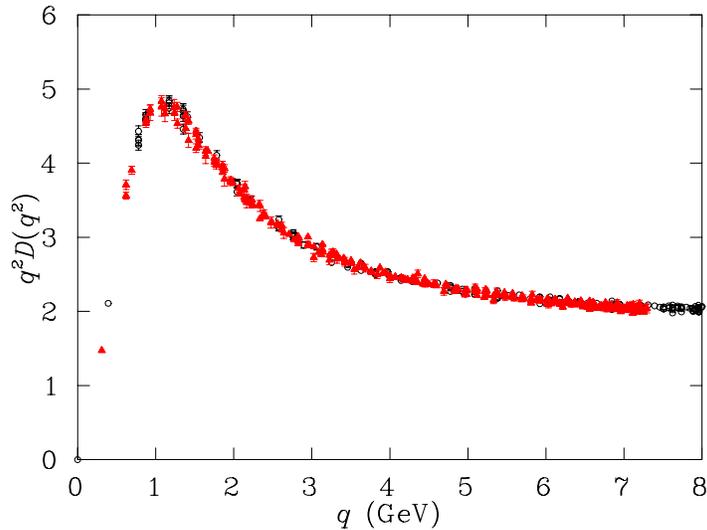,angle=90,width=11cm}
\end{center}
\caption{The momentum-enhanced propagator in \lapI\ gauge for the Wilson gauge 
action at $\beta = 6.0$ (open circles; data from Ref.~\cite{Ale01}) and the 
improved gauge action at $\beta = 4.60$ (filled triangles).  Both lattices are 
$16^3\times 32$.  Data has been cylinder cut. The two are in excellent 
agreement.  In Ref.~\cite{Bon01} the difference in normalization between the
Wilson and improved actions was seen to be $\sim 1.08$.  After taking this 
into account, the relative normalization here is 
$Z_3(\beta=4.60) = 1.07 Z_3(\beta=6.0)$}.
\label{fig:lap1_usandthem_polcut}
\end{figure}

We start by checking that our finest lattice, at $a=0.125$ fm, is 
``fine enough'', by comparing the propagator with that of Alexandrou 
\etal~\cite{Ale01}.  Fig.~\ref{fig:lap1_usandthem_polcut}
shows the momentum-enhanced propagator, $q^2 \, D(q^2)$, in \lapI\ gauge for 
ensemble 2i (improved action) 
compared to the data from Ref.~\cite{Ale01} for the Wilson action at 
$\beta=6.0$\footnote{For best comparison with our data, we have used 
$a^{-1} = 2.0$ GeV for $\beta=6.0$.}.  The two are in excellent agreement.

As the gluon propagator has been extensively studied in Landau gauge, it makes
sense to understand the Laplacian gauge propagator by comparing it to
that in Landau gauge.  In accordance with custom, we will
discuss the momentum-enhanced propagator, $q^2 D(q)$.  We define the relative 
$Z_3$ renormalization constant
\begin{equation}
Z_R \equiv \frac{Z_3(\text{Landau})}{Z_3(\text{Laplacian})},
\end{equation}
and choose to perform this matching at $\mu = 4.0$ GeV.
The purpose of this is simply to make the (bare) gluon propagators agree in
the ultraviolet for easy comparison between the gauges. 

\begin{figure}[h]
\begin{center}
\epsfig{figure=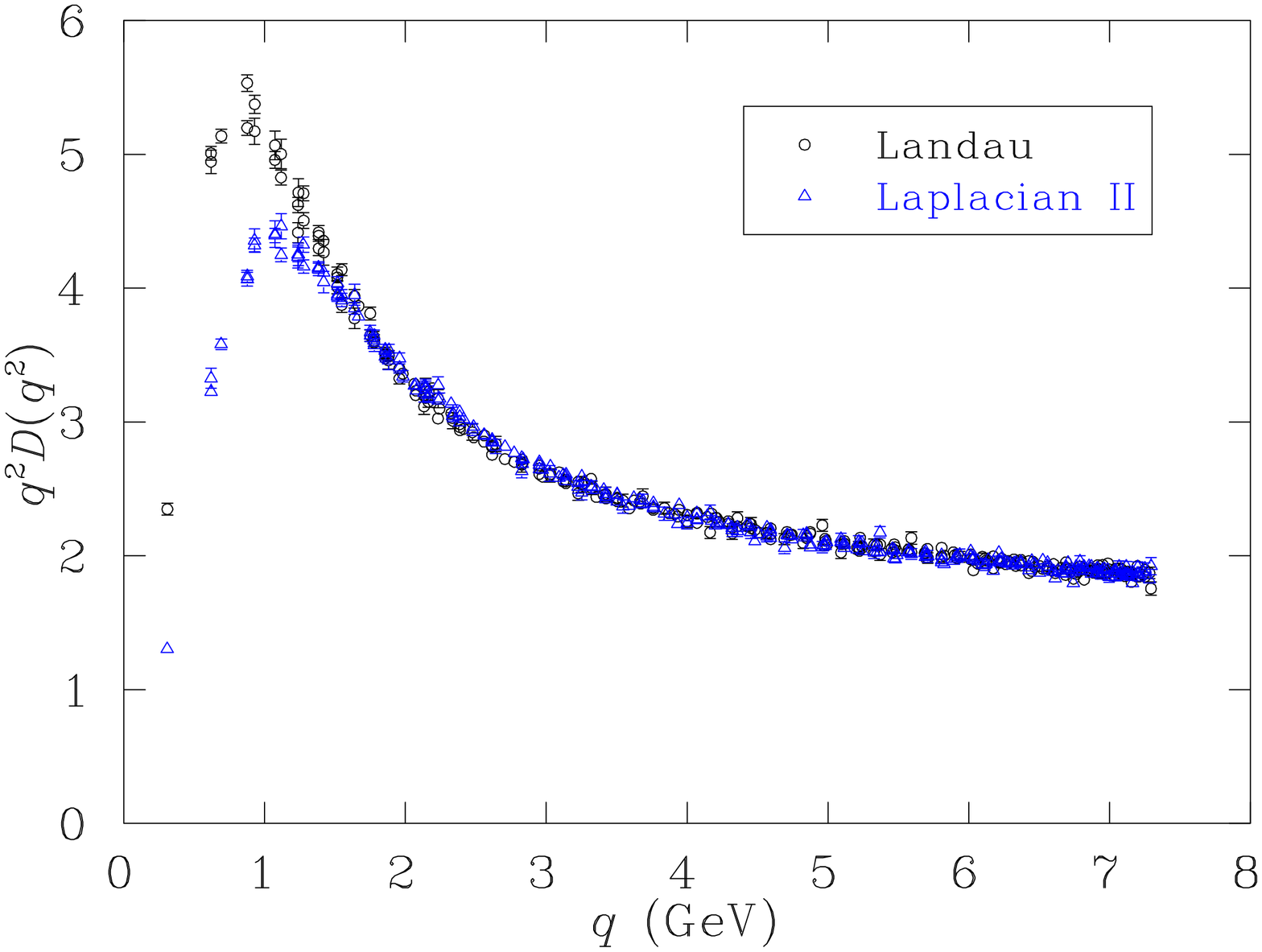,angle=90,width=11cm}
\end{center}
\begin{center}
\epsfig{figure=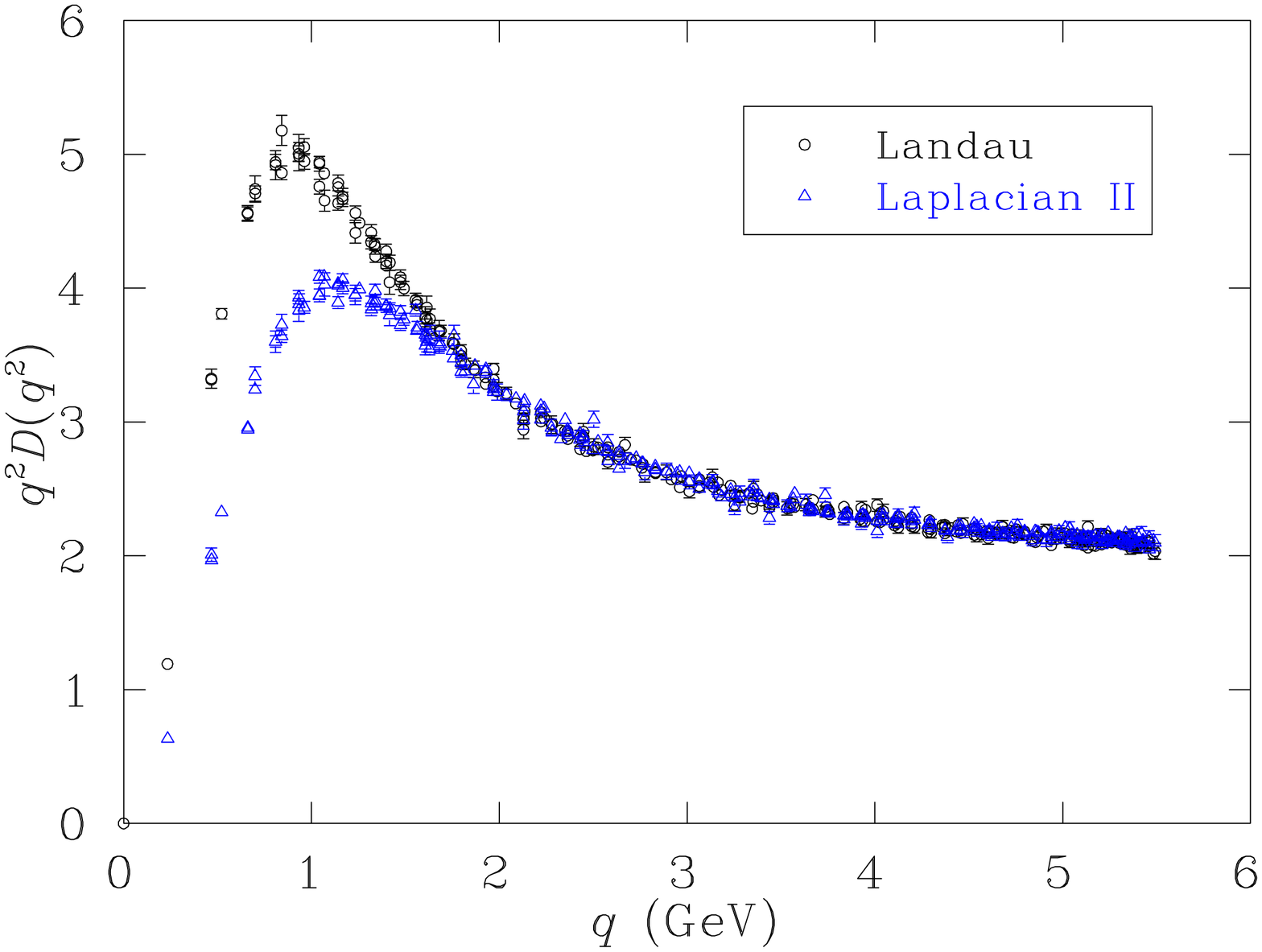,angle=90,width=11cm}
\end{center}
\caption{Comparison of the gluon propagator in Landau and \lapII\ gauges for
lattice 2i (top) and 3 (bottom) ($16^3 \times 32$, improved action, 
$\beta=4.60$ and 4.38 respectively).  Data has been
cylinder cut.  The relative normalizations are $Z_R(\beta=4.60) = 1.075$
and $Z_R(\beta=4.38) = 1.20$.}
\label{fig:comp_b460_cut}
\end{figure}

We show the gluon propagator in both Landau and \lapII\ gauges in 
Fig.~\ref{fig:comp_b460_cut}.
In this figure, the data is for the largest finer lattices, 2i and 3.  The 
data has been cylinder cut~\cite{Lei99,Bon01} to make comparison easier. 
As was seen in Ref.~\cite{Ale01, Ale02}, the gluon propagator in Laplacian 
gauge looks very similar to the Landau gauge case
matching up in the ultraviolet but with a somewhat lower infrared hump.  

Having compared the gluon propagator in \lapII\ gauge to Landau gauge, we now 
compare it to other implementations of the Laplacian gauge. 
We expect each implementation
to provide a well defined, unambiguous, but different gauge.  As we saw when
comparing Landau and \lapII\ gauges, there is some difference in normalization
between the propagators in the different gauges, so we define, again at 
$\mu = 4.0$ GeV
\begin{equation}
Z_\del = \frac{Z_3(\lapII)}{Z_3(\lapI)}.
\end{equation}

\begin{figure}[h]
\begin{center}
\epsfig{figure=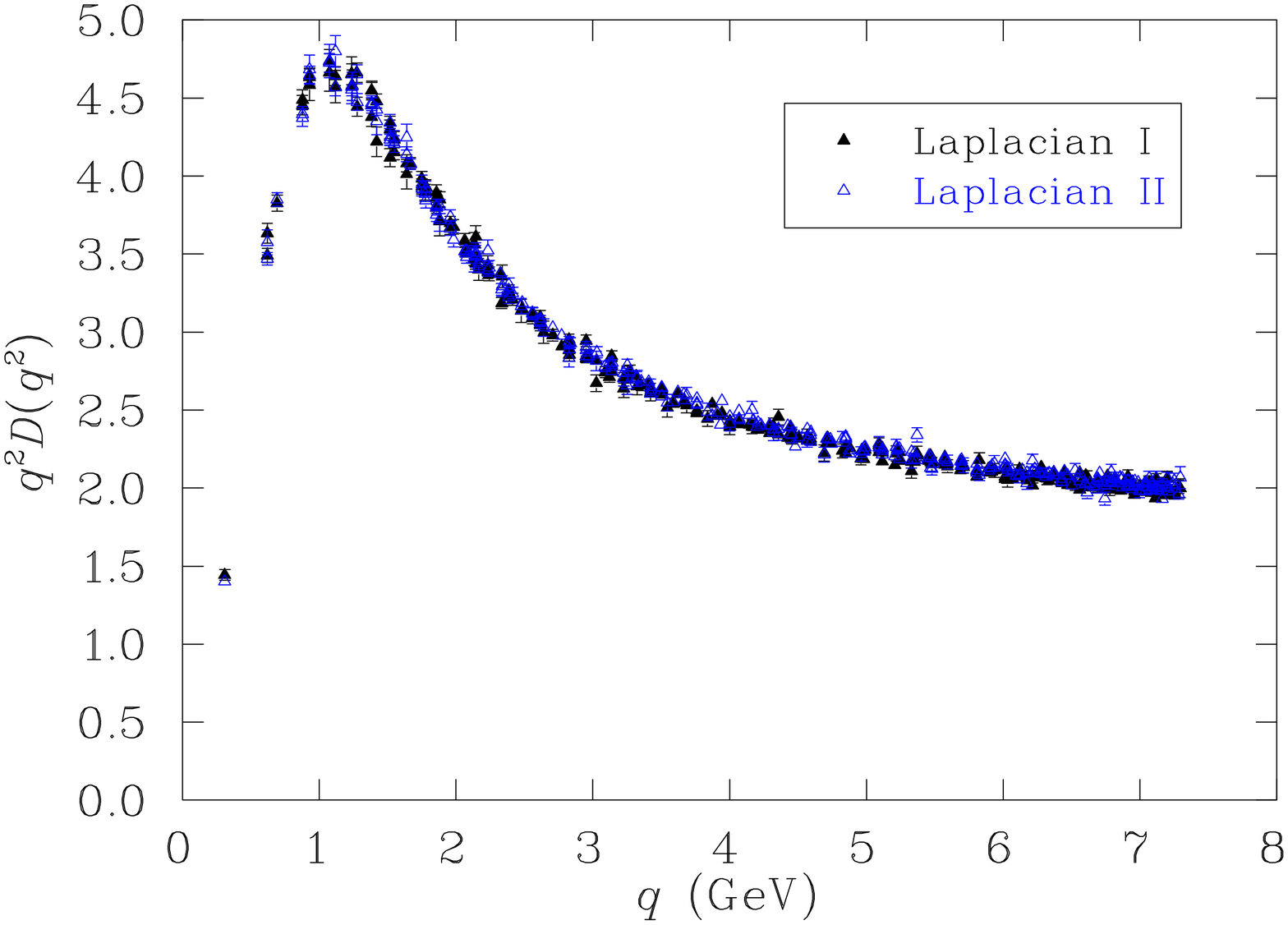,angle=90,width=11cm}
\end{center}
\caption{The momentum-enhanced propagator from ensemble 2i
($\beta=4.60, 16^3 \times 32$, improved action) in \lapI\ and \lapII\
gauges.  Data has been cylinder cut. The two gauges produce nearly identical
gluon propagators, up to a small relative normalization ($Z_\del = 0.98$).}
\label{fig:lap12_b460_cut}
\end{figure}

In Fig.~\ref{fig:lap12_b460_cut}, the momentum-enhanced propagator is plotted 
in \lapI\ and \lapII\ gauges for one of the fine lattices (ensemble 2i).  There
is a small 
relative normalization ($Z_\del = 0.98$), but otherwise there is no
significant difference between them, neither in the ultraviolet nor the 
infrared.   \lapI\ and \lapII\ also show comparable
performance in terms of rotational symmetry.

\begin{figure}[h]
\begin{center}
\epsfig{figure=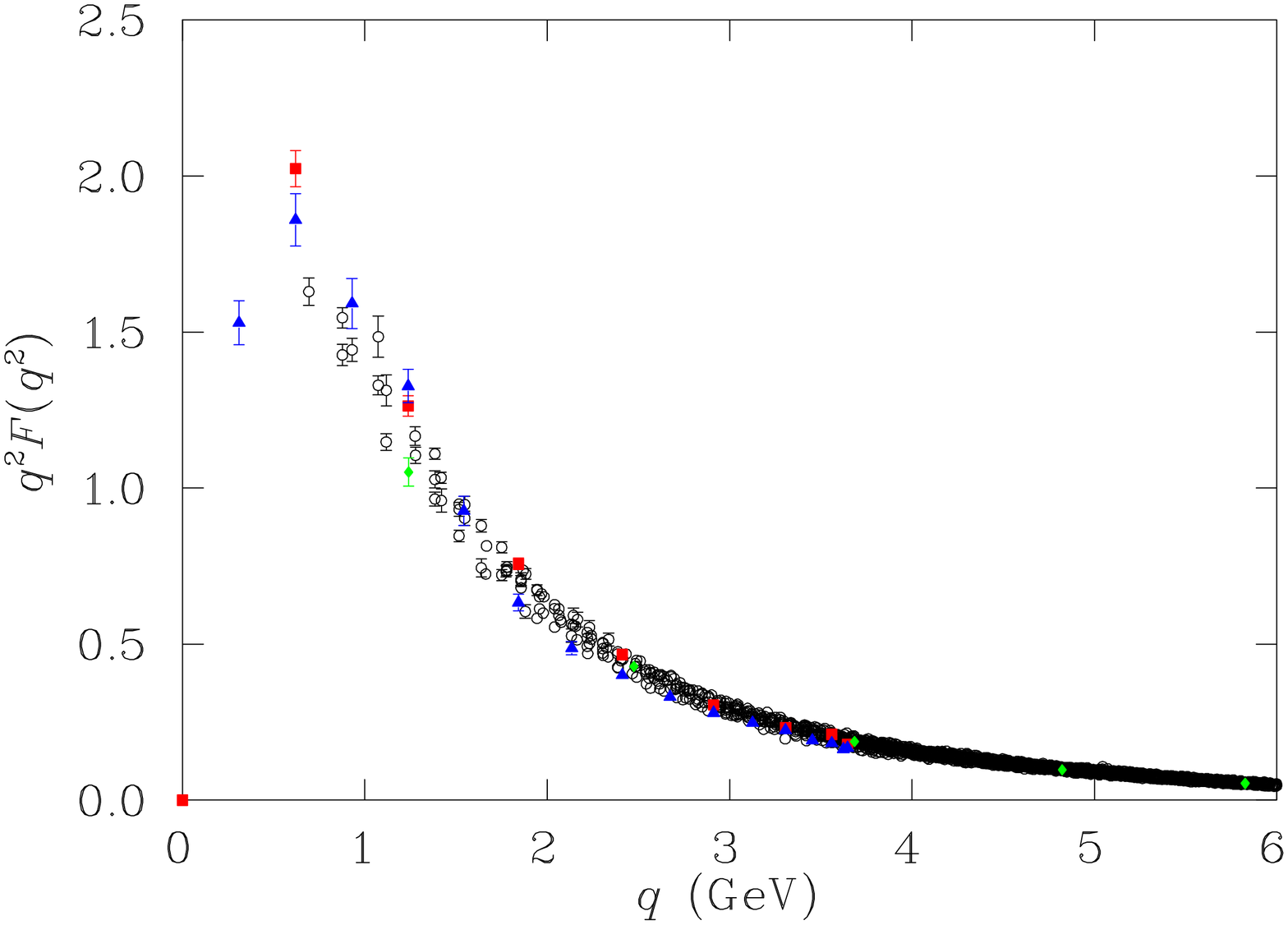,angle=90,width=11cm}
\end{center}
\begin{center}
\epsfig{figure=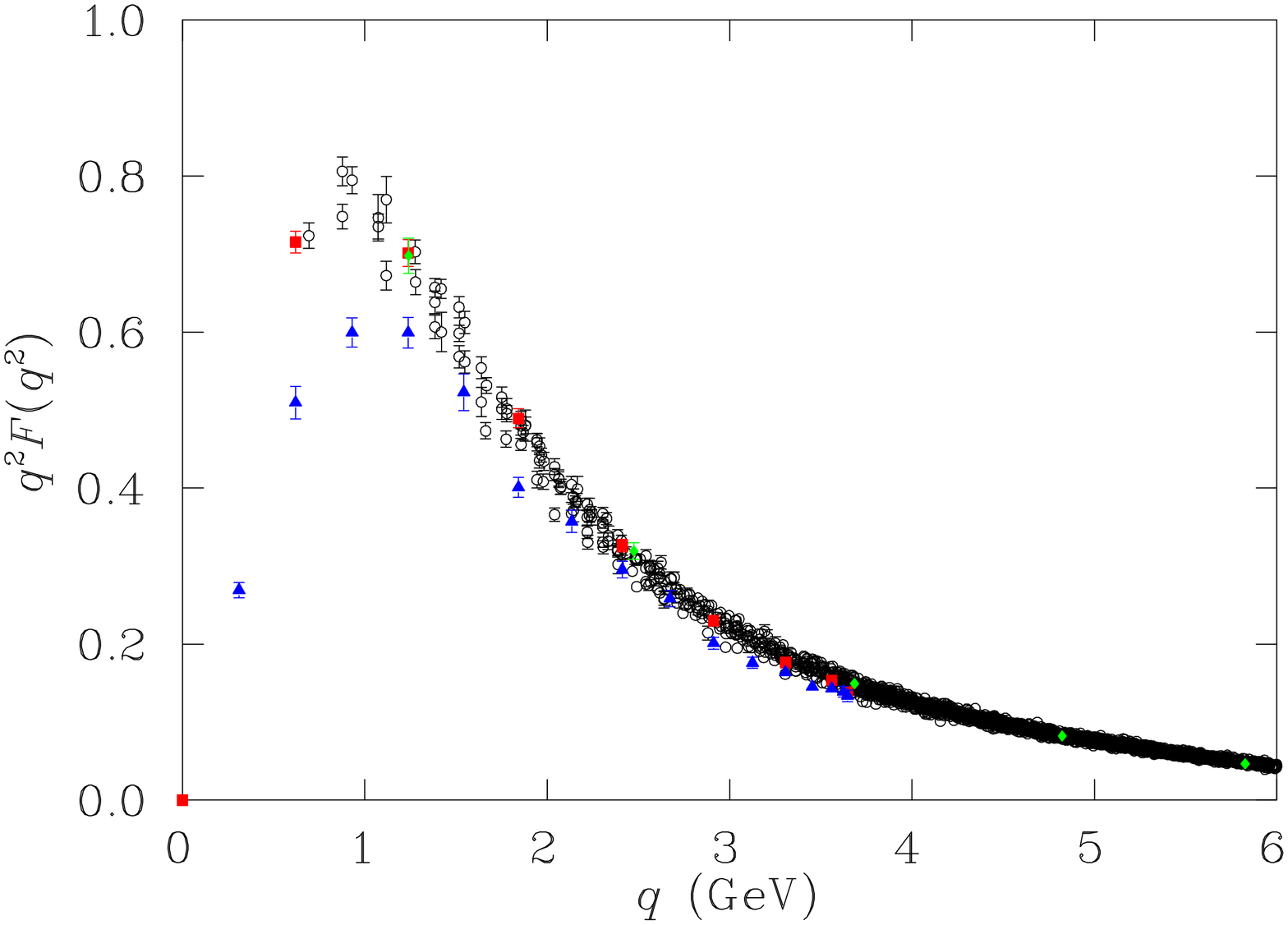,angle=90,width=11cm}
\end{center}
\caption{Comparison of the longitudinal part of the gluon propagator in 
\lapI\ (above) and \lapII\ (below) gauges for
lattice 2i ($\beta=4.60, 16^3 \times 32$, improved action).  There have been
no data cuts or renormalization.  Note that the vertical scales are different, 
so the longitudinal component is smaller in \lapII\ gauge than \lapI.  The data
has been sorted into points where all the momentum is in the temporal 
direction, spatial cartestian direction, four-diagonal, and the rest.  The 
separation of temporal points (triangles) from spatial cartesian points 
(squares) suggests that this part of the gluon propagator has stronger volume
dependence in \lapII\ gauge than in \lapI\ gauge.  In a
standard covariant gauge this would be a constant, $\xi$ (zero, for Landau
gauge). }
\label{fig:comp_b460_long_all}
\end{figure}

One difference between Landau and Laplacian gauge is that in the former, the
gluon propagator has no longitudinal component.  We see in 
Fig.~\ref{fig:comp_b460_long_all} that the longitudinal part of the propagator
does indeed vanish in the ultraviolet, which is consistent with 
approaching Landau gauge, but gains strength in the infrared.  Comparing 
\lapI\ and \lapII\ gauges we note that while the 
transverse parts look alike, the longitudinal behavior is quite distinct.  
The separation of squares and triangles in \lapII\ gauge suggests that
$F(q^2)$ has stronger volume dependence in that gauge.  
For a comparison between Landau, \lapI\ and \lapII\ gauges for the quark 
propagator see Ref.~\cite{quarkprop}.

\begin{figure}[p]
\begin{center}
\epsfig{figure=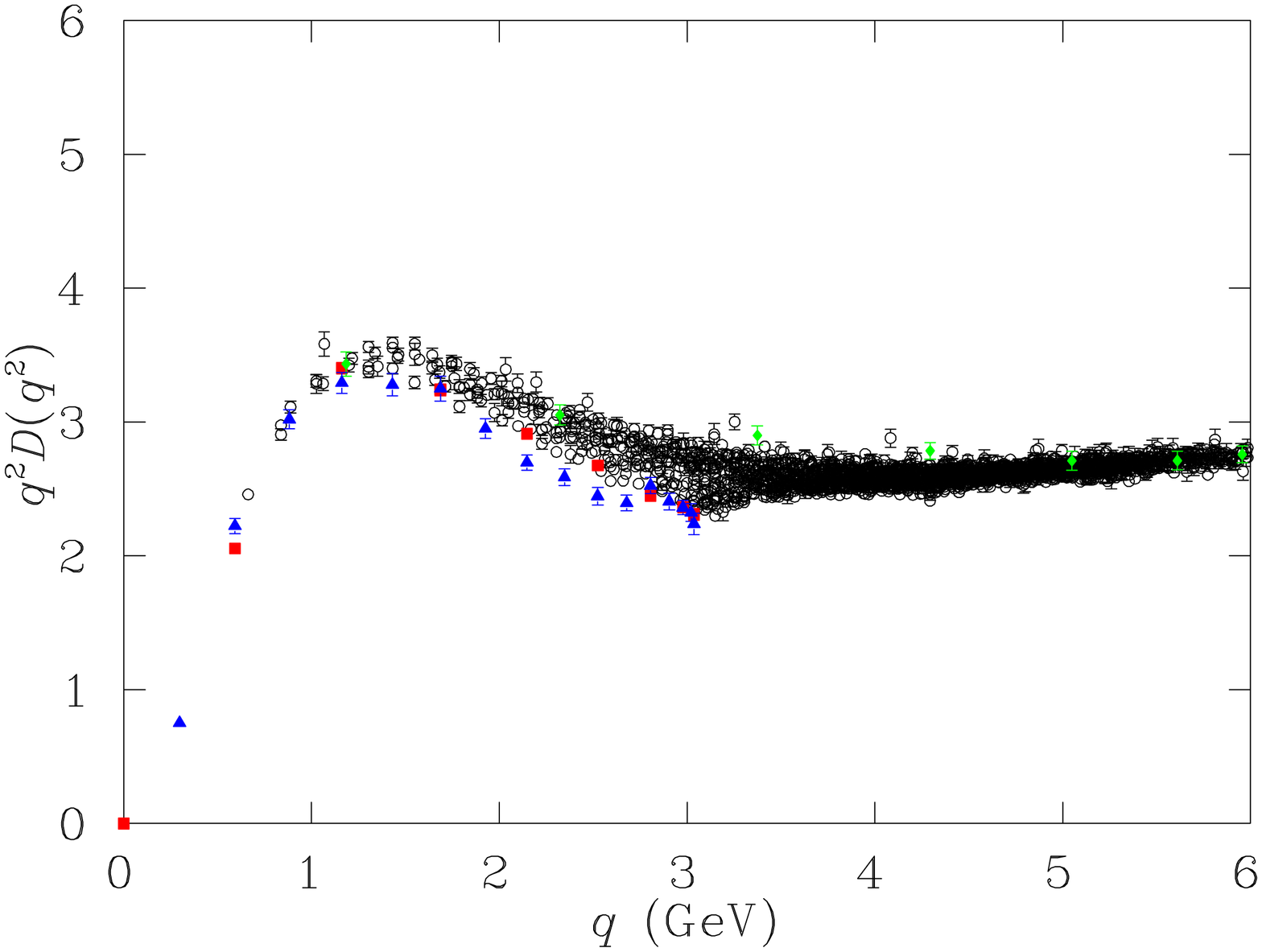,angle=90,width=11cm}
\end{center}
\caption{momentum-enhanced propagator from 76 configurations in \lapIII\ gauge 
from 
lattice 1w ($\beta=5.85, 16^3 \times 32$, Wilson action).  There have been
no data cuts.  This is clearly a very bad gauge fixing.}
\label{fig:lap3_b585_all}
\end{figure}

\lapIII\ gauge works badly, failing even to reproduce the correct asymptotic 
behavior.  Fig.~\ref{fig:lap3_b585_all} shows data from only 76 configurations 
as the gauge fixing failed entirely on four of them.  In $SU(3)$, the polar
decomposition involves calculating determinants which, to our numerical 
precision, can become vanishingly small, in some cases even turning negative
at some sites.
\lapIII\ gauge was also seen to be a very poor gauge for studying the quark
propagator~\cite{quarkprop}.

\subsection{Coarser lattices}

When comparing results from ensembles with different simulation parameters
we need to consider three possible effects:

\begin{enumerate}

\item The dependence of $Z_3$ on the lattice spacing,

\item Errors due to the finite lattice spacing, especially when probing
	momenta near the cutoff,

\item Finite volume effects, especially in the infrared.

\end{enumerate}

\begin{figure}[h]
\begin{center}
\epsfig{figure=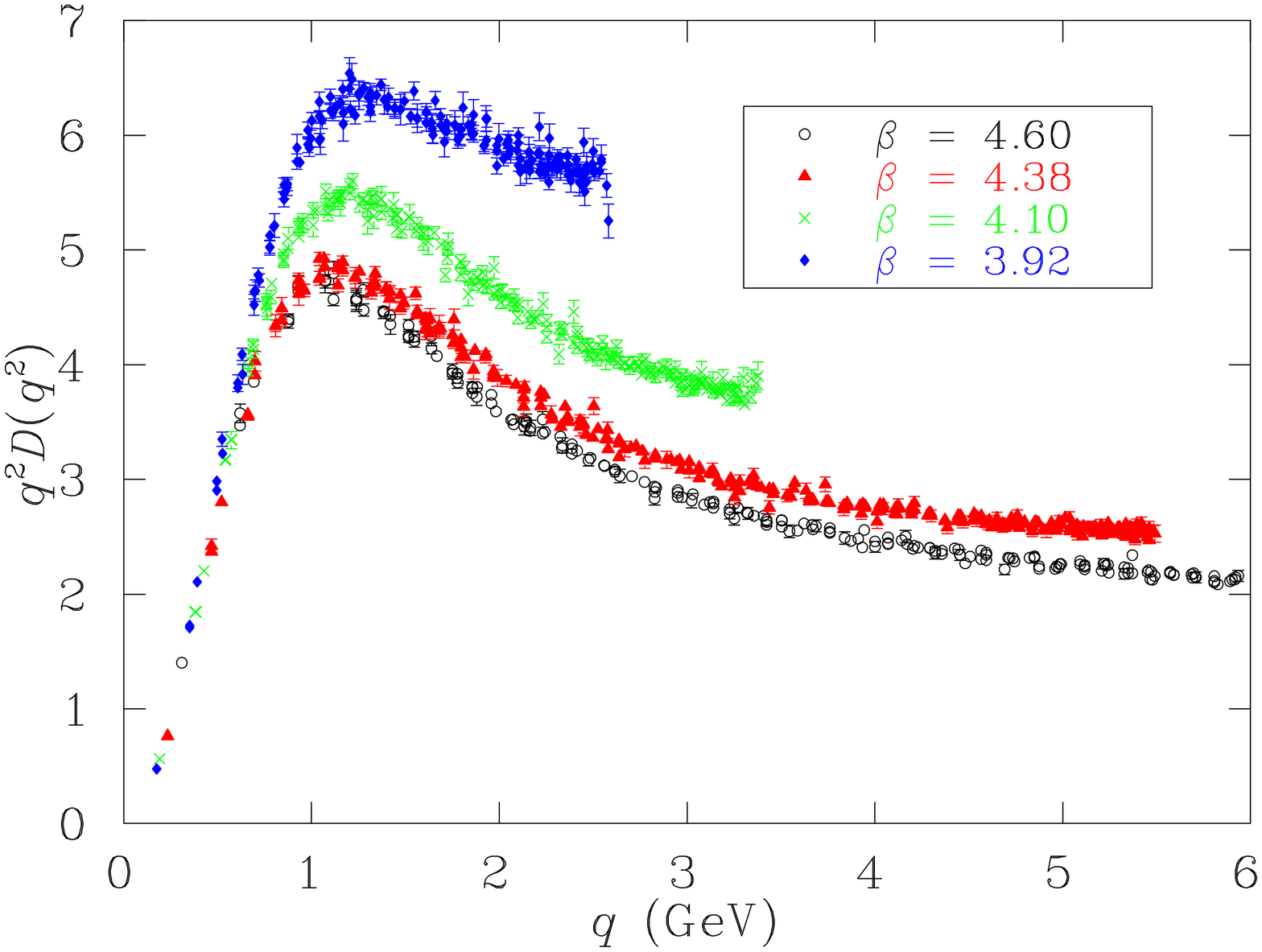,angle=90,width=11cm}
\end{center}
\caption{The momentum-enhanced propagator in \lapII\ gauge at 
$\beta = 4.60$, 4.38, 4.10 and 3.92 (ensembles 2i, 3, 4 \& 5).  Data has been 
cylinder cut.  This figure shows the large sensitivity of $Z_3$ to lattice 
spacing in Laplacian gauge, quite unlike Landau gauge.}
\label{fig:lap2_comp_cut}
\end{figure}

In Landau gauge, the dependence of the gluon 
propagator renormalization, $Z_3(a)$, on the cutoff is very weak.
$Z_3$ is approximately constant with respect to the lattice 
spacing~\cite{Bon01}.  In this case it is easy to compare propagators 
produced on a wide range of lattice spacings.  
In Fig.~\ref{fig:lap2_comp_cut} we plot the momentum-enhanced propagator on four 
lattices, which have $a = 0.125$, 0.166, 0.270 and 0.353 fm.  We see a very
different situation to the one observed in Landau gauge~\cite{Bon01}: the
propagators appear to agree in the the deep infrared, yet diverge as the
momentum increases.  The difference is small for the two finest lattices -
and non-existent in Fig.~\ref{fig:lap1_usandthem_polcut} - but
quite dramatic for the coarsest.  

\begin{figure}[h]
\begin{center}
\epsfig{figure=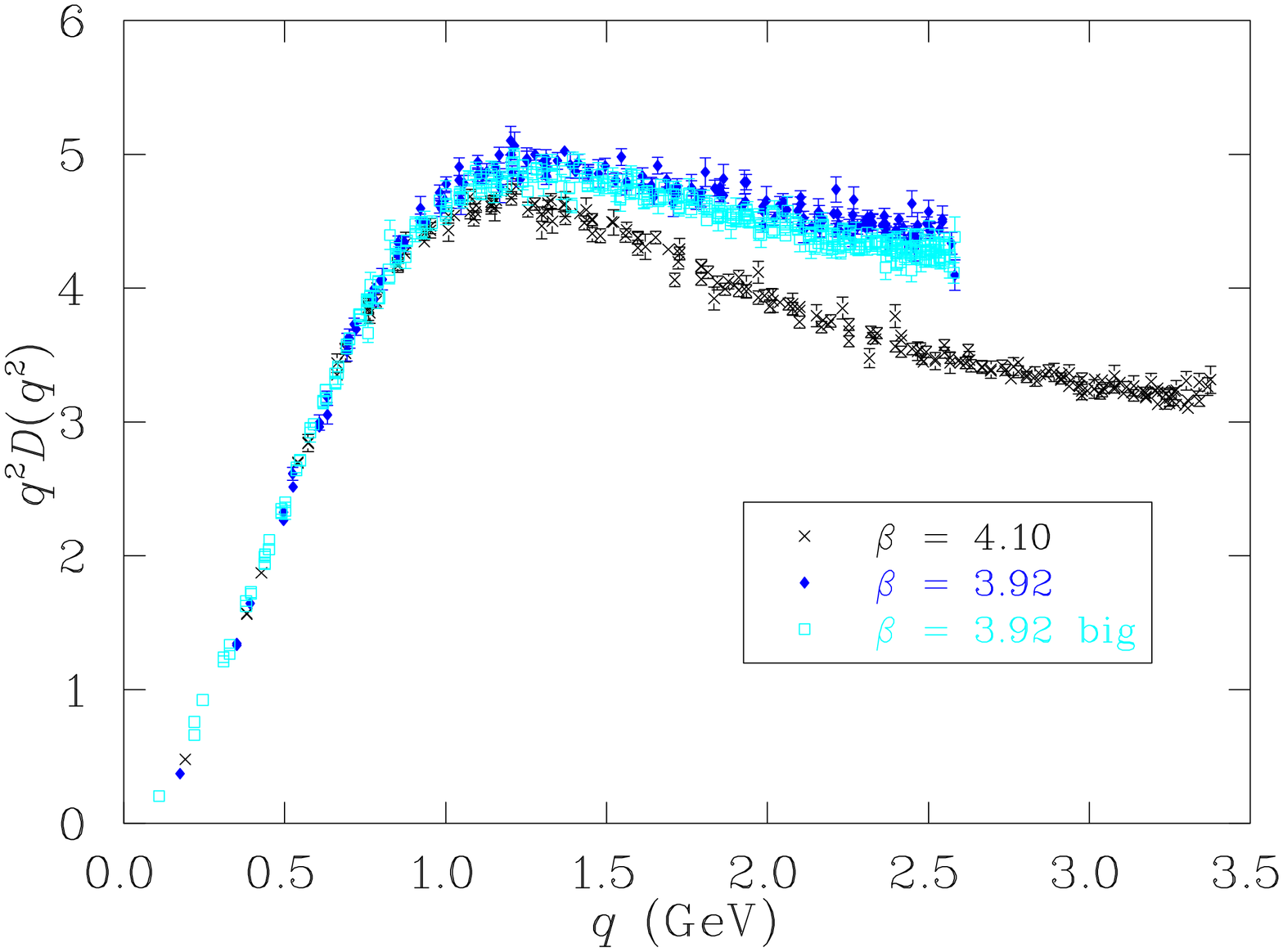,angle=90,width=11cm}
\end{center}
\caption{The momentum-enhanced propagator from ensembles 3, 4 and 5 in 
\lapII\ gauge.  Data has been cylinder cut and normalized at 0.6 GeV.}
\label{fig:lap2_coarse_normpol}
\end{figure}

The correct way to compare the propagators is to normalize them at some common,
``safe'' momentum, i.e., one where we expect finite lattice spacing and finite
volume effects to both be small.  We choose $\mu = 0.6$ GeV and show the 
results in Fig.~\ref{fig:lap2_coarse_normpol}.  
Multiplying the propagator by $q^2$ in constructing the 
momentum-enhanced propagator
$q^2 \, D(q^2)$ reveals a rapid divergence in the ultraviolet.
Yet normalizing at higher momenta results in data
sets that agree nowhere except for at the scale $\mu$.  It is interesting 
that the 
propagators from ensembles 5 and 6, which have the same lattice spacing, have
slightly different normalizations.

\begin{figure}[h]
\begin{center}
\epsfig{figure=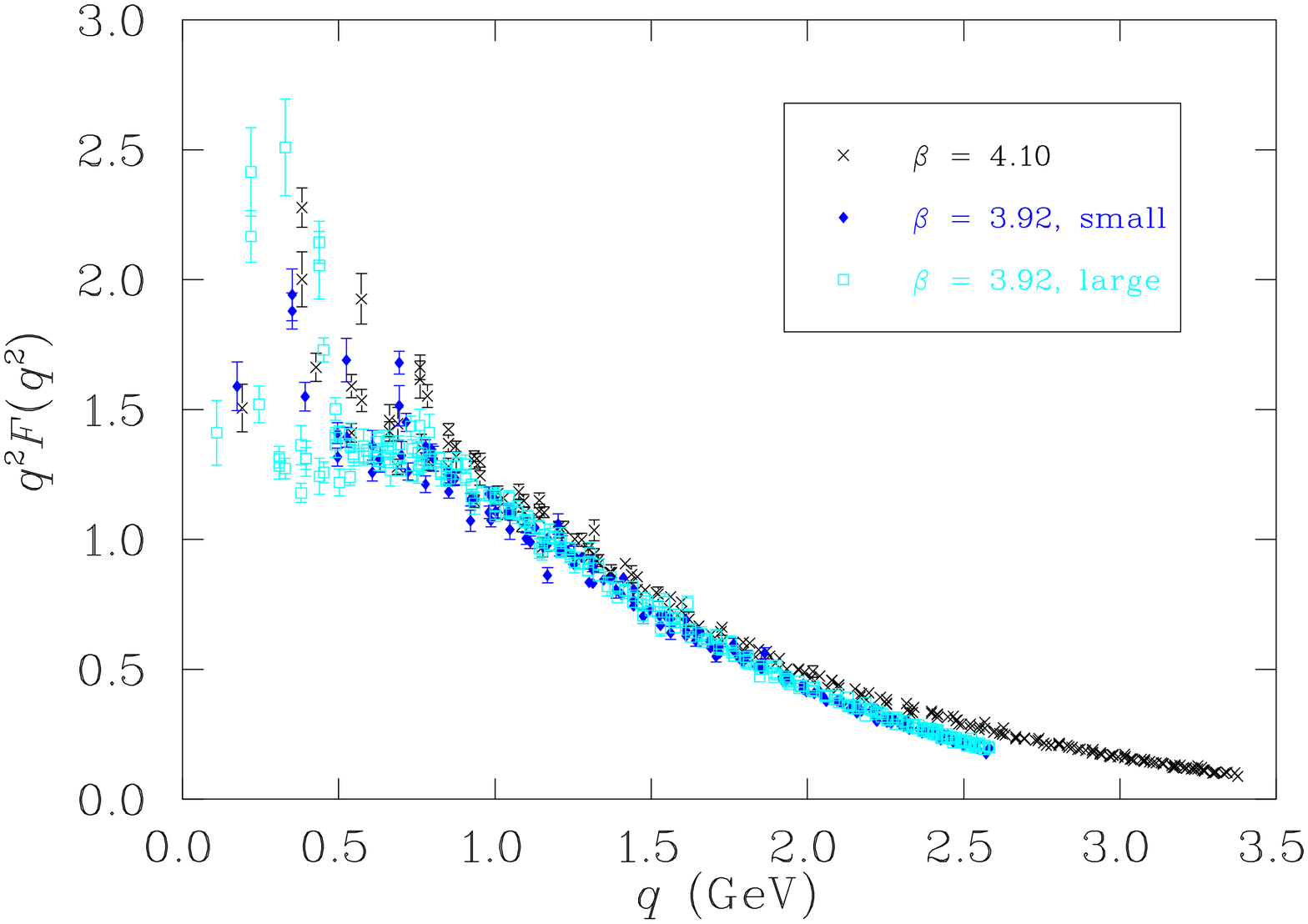,angle=90,width=11cm}
\end{center}
\begin{center}
\epsfig{figure=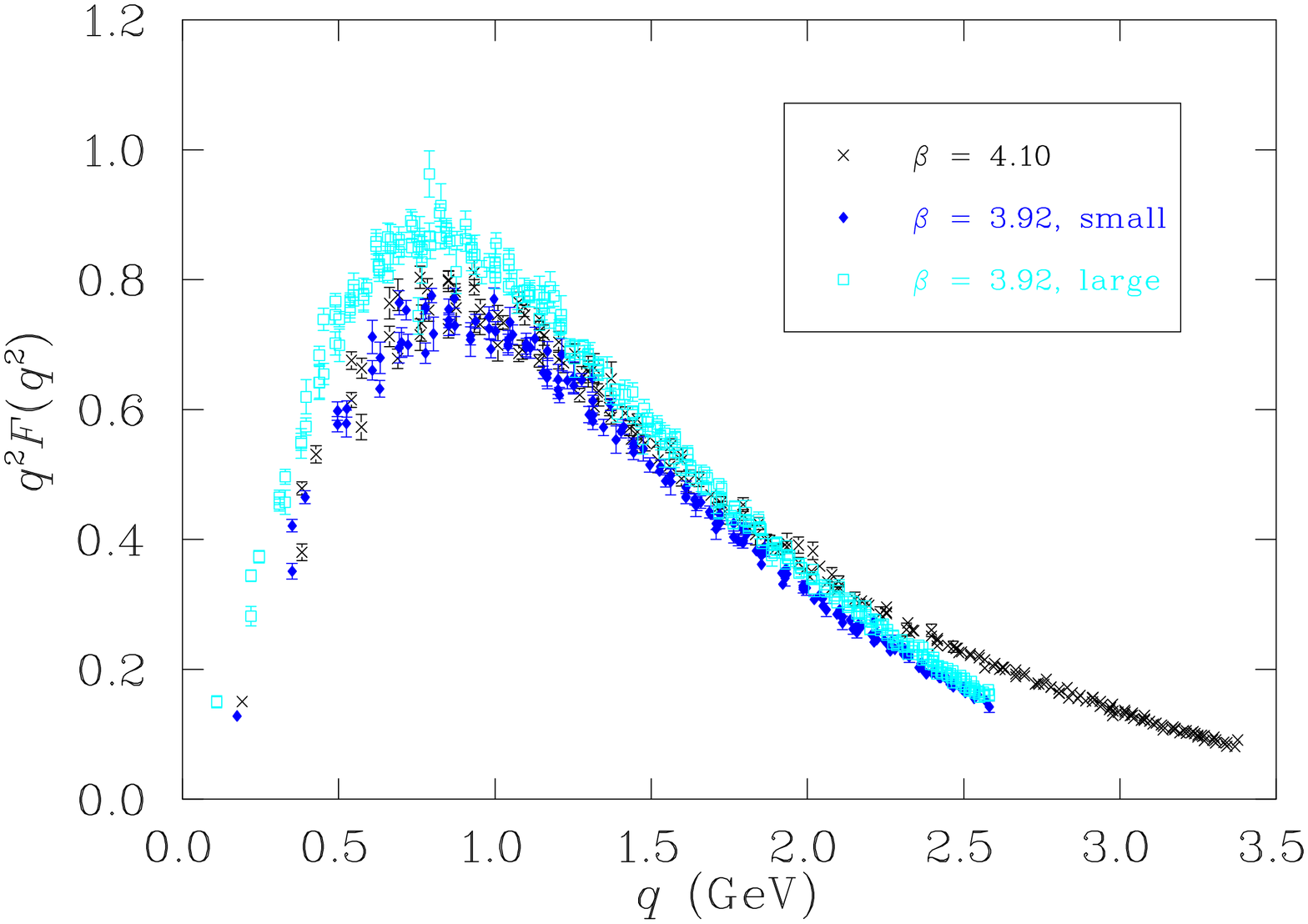,angle=90,width=11cm}
\end{center}
\caption{Comparison of the longitudinal part of the gluon propagator in 
\lapI\ (top) and \lapII\ (bottom) gauges for
ensembles 4, 5 \& 6.  Data has been cylinder cut.  The small and large 
$\beta = 3.92$ lattices give diverging results even at large momenta in the
\lapII\ gauge.}
\label{fig:comp_coarse_long}
\end{figure}

Also, as the lattice spacing is increased, the relative normalization between 
the gluon propagators in \lapI\ and \lapII\ gauges slowly diverges from one.  
As was seen above (Fig.~\ref{fig:comp_b460_long_all}), these two Laplacian 
gauges produce gluon propagators with rather different longitudinal components.
The longitudinal part of the gluon propagator, multiplied by $q^2$, is plotted 
in \lapI\ and \lapII\
gauges for ensembles 4-6 in Fig.~\ref{fig:comp_coarse_long}, using the same
normalization determined for the transverse parts.  Interestingly, 
the longitudinal part appears
to be more affected by the finite volume of the lattice than the transverse
part.  In \lapII\ gauge $q^2F(q^2)$ may
return to zero as $q^2 \rightarrow 0$, while in \lapI\ gauge a small, 
non-zero value appears likely, however, more work is required.

In previous studies~\cite{Ale01,Ale02} it was observed that the infrared 
gluon propagator saturates at a small volume ($\sim 1 \text{ fm}^4$). 
To further explore this, we also study the propagator at zero four-momentum.  
As previously discussed,
only the full propagator, $\mathcal{D}(0)$, can be calculated at zero 
four-momentum.  In order to compare results on all of our
lattices we normalize the data at 1 GeV.  This represents a compromise and
is certainly not ideal for all the data sets, hence there is some systematic
error.  These results are shown in Table~\ref{tab:zeromom}.

\begin{table}[bt]
\centering
\begin{ruledtabular}
\begin{tabular}{ccccccc}
   & Dimensions	     & $\beta$ & $a$ (fm) & Volume $\text{(fm}^4\text{)}$ &
$\mathcal{D}(0)$ - \lapI\ & $\mathcal{D}(0)$ - \lapII  \\
\hline
1  & $12^3\times 24$ &   4.60  &   0.125  & 10.1 & 16.6(5) & 16.3(4) \\
2i & $16^3\times 32$ &   4.60  &   0.125  & 32.0 & 17.1(5) & 16.0(4) \\
3  & $16^3\times 32$ &   4.38  &   0.166  & 99.5 & 16.7(6) & 14.1(3) \\
4  & $12^3\times 24$ &   4.10  &   0.270  & 220  & 19.0(8) & 14.6(2) \\
5  & $10^3\times 20$ &   3.92  &   0.353  & 300  & 20.8(9) & 14.6(4) \\
6  & $16^3\times 32$ &   3.92  &   0.353  & 2040 & 52(5)   & 35(2)
\end{tabular}
\end{ruledtabular}
\caption{\label{tab:zeromom}The full propagator at zero four-momentum  
in \lapI\ and \lapII\ gauges, normalized at 1 GeV for comparison.}
\end{table}

If we restrict ourselves to ensembles 1-5, we see that, given the uncertainties
discussed, the sensitivity of $\mathcal{D}(0)$ to volume is indeed small.  We
also see the trend, already noted above, for \lapI\ and \lapII\ gauges to 
become more different as the lattice spacing increases, another example of the
discretization sensitivity of the Laplacian gauge.  In the case of the
very large lattice, ensemble 6, this sensitivity has become extreme.

\begin{figure}[h]
\begin{center}
\epsfig{figure=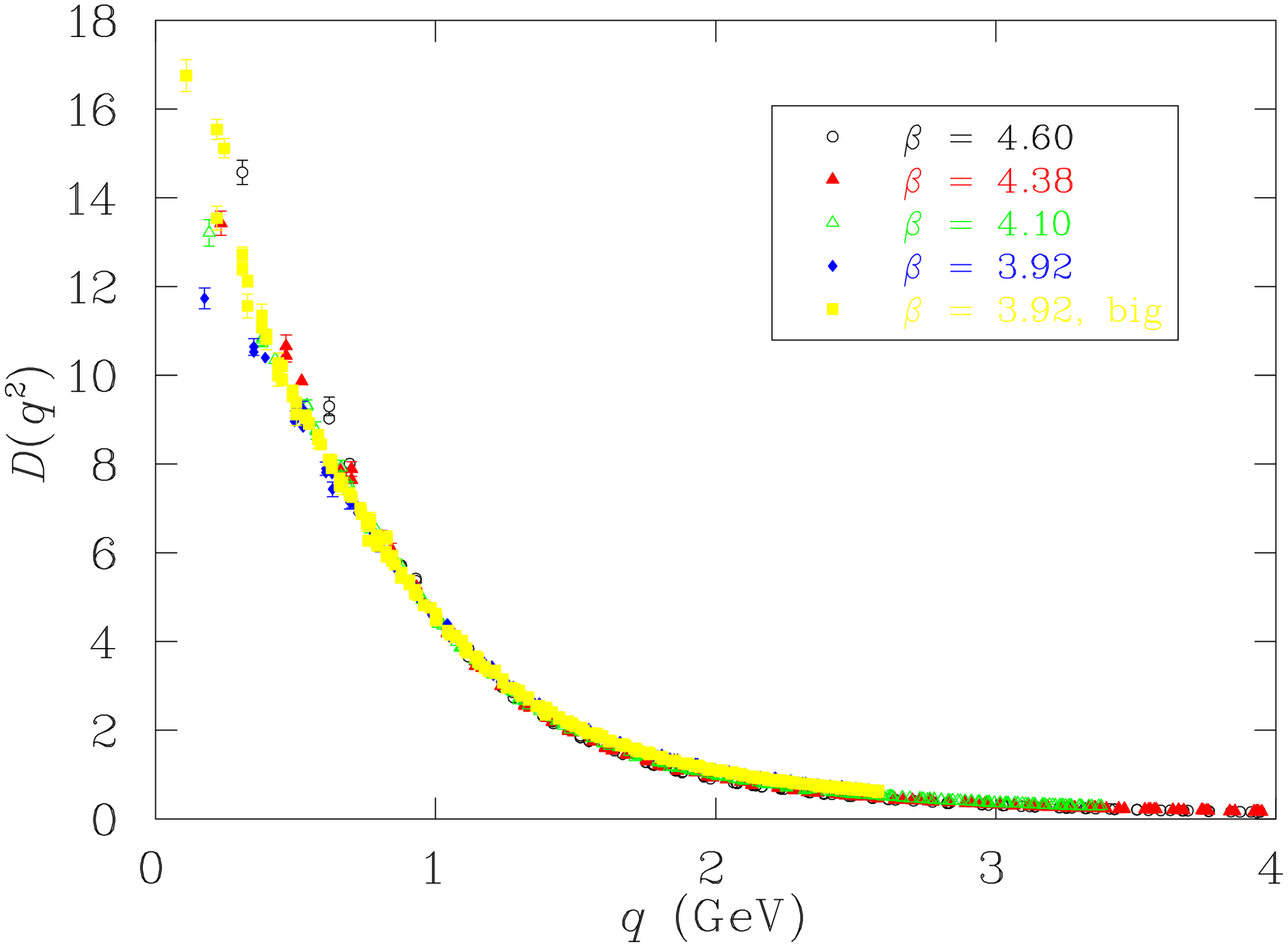,angle=90,width=11cm}
\end{center}
\caption{Comparison of the transverse part of the gluon propagator in 
\lapII\ gauge for ensembles 1i, 2-6.  Data has been cylinder cut.  
The propagators have been normalized at 1 GeV.}
\label{fig:lap2_all_prop}
\end{figure}

We examine this another way through the transverse propagator, shown in
Fig.~\ref{fig:lap2_all_prop} for ensembles 2i, 3 - 6.
The data here corresponds to Table~\ref{tab:zeromom}, having been normalized 
at 1 GeV.  The propagators are consistent down to low
momenta, $\sim 200$ MeV, where we strike trouble.  Not only is there a spread
between the data sets, but for ensemble 6, the point from purely temporal 
momentum is much lower than that from spatial momentum.  The situation is the
same in \lapI\ gauge.

\section{Conclusions}

We have made a detailed study of the gluon propagator on coarse lattices with
an improved action in Laplacian gauges.  We have described and tested three
implementations of Laplacian gauge.  \lapI\ (QR decomposition) and
\lapII\ (Maximum trace) gauges produce similar 
results for the scalar transverse gluon propagator, but rather different 
longitudinal components.  \lapIII\, for numerical reasons, works very poorly 
in $SU(3)$.  

At sufficiently small lattice spacing, the transverse
part of the gluon propagator is very similar in Laplacian gauge to that in 
Landau gauge.  Laplacian gauge, however, exhibits great sensitivity to the
lattice spacing, making results gained from coarse lattices difficult to 
compare with those from finer lattices.  This is very different to the 
situation in Landau gauge.  By comparing the coarse data sets at sufficiently
low momentum, however, we have seen a great deal of consistency.  In the 
deep infrared, the results from the largest lattice are difficult to reconcile 
with the other data.  Excluding that lattice, the total propagator shows 
little sign of volume dependence.  
The most likely explanation of the (unimproved) Laplacian gauge results seen
here is that on our coarsest lattices (improved lattices with $\beta = 3.92$, 
4.10, and to some extent even $\beta = 4.38$), the lattice artifacts are much 
more severe than in the improved Landau gauge case.  On the very coarse 
$\beta = 3.92$ and 4.10 lattices, it seems very likely that finite volume and 
discretization errors are actually being coupled together by the 
Laplacian gauge fixing.  By implementing an improved Laplacian gauge fixing on 
these lattices we anticipate that these errors will decouple on these lattices
and we will be in a better position to estimate the infinite volume and 
continuum limits of the different implementations of Laplacian gauge-fixing.
Further studies, including an improved
Laplacian gauge fixing, will hopefully clarify these issues.

\begin{acknowledgments}
The authors would like to thank Constantia Alexandrou for helpful discussions
during the Lattice Hadron Physics workshop in Cairns, Australia.
Computational resources of the Australian National Computing Facility
for Lattice Gauge Theory are gratefully acknowledged.
The work of UMH and POB was supported in part by DOE contract 
DE-FG02-97ER41022.  DBL and AGW acknowledge financial support from the
Australian Research Council.
\end{acknowledgments}



\begin{thebibliography}{99}


\bibitem{background} Jeffrey E.~Mandula, \emph{The Gluon Propagator},\/
	hep-lat/9907020; 
	L.~von Smekal and R.~Alkofer,
	\emph{What the Infrared Behavior of QCD Green Functions can tell us
	about Confinement in the Covariant Gauge},
	hep-ph/0009219;
	For a recent demonstration of the connection between the gluon
	propagator and confinement, see e.g., Kurt Langfeld,
	hep-lat/0204025.

\bibitem{Bec99} D.~Becirevic \emph{et al.}, 
	\rf{\pr}{D 60}{094509}{1999};
	D.~Becirevic \emph{et al.},
   	\rf{\pr}{D 61}{114508}{2000}.

\bibitem{models}  See, for example,
	C.~D.~Roberts and A.~G.~Williams,
        \rf{Prog.\ Part.\ Nucl.\ Phys.\ } {33}{477}{1994}.

\bibitem{Cuc01} A.~Cucchieri and D.~Zwanziger,
	\npps{106}{694}{2002}.

\bibitem{Giu01} L.~Giusti \emph{et al.},
	\npps{106}{995}{2002}.

\bibitem{Lei99} D.B.~Leinweber, J-I.~Skullerud and A.G.~Williams,
	\rf{\pr}{D 60}{094507}{1999}; 
	\rf{Erratum-ibid.}{D 61}{079901}{2000}.

\bibitem{Bon00} F.D.R.~Bonnet, P.O.~Bowman, D.B.~Leinweber \&
	A.G.~Williams, 
	\rf{\pr}{D 62}{051501}{2000}.

\bibitem{Bon01} F.D.R.\ Bonnet, P.O.\ Bowman, D.B.\ Leinweber, 
	A.G.\ Williams and J.M.\ Zanotti,
	\rf{\pr}{D 64}{034501}{2001}.

\bibitem{Gri78} V.N.~Gribov, \rf{\np}{B 139}{1}{1978}.

\bibitem{Het98} J.E.~Hetrick \& Ph.~de Forcrand,  
	\npps{63}{838}{1998}.
 
\bibitem{evolutionary} J.F.~Markham \& T.D.~Kieu, 
	\npps{73}{868}{1999}; O.~Oliveira and P.J.~Silva,
	\npps{106}{1088}{2002}.

\bibitem{Wil02} A.G.~Williams,
	\npps{109}{141}{2002}.

\bibitem{Vin92} J.C.~Vink and U-J.~Wiese, 
	\rf{\physl}{B 289}{122}{1992}.

\bibitem{Vin95} J.C.~Vink,
 	\rf{\pr}{D 51}{1292}{1995}.	

\bibitem{Ale01} C.~Alexandrou, Ph.~de Forcrand and E.~Follana,
	\rf{\pr}{D 63}{094504}{2001}.

\bibitem{Ale02} C.~Alexandrou, Ph.~de Forcrand and E.~Follana,
	hep-lat/0112043. 

\bibitem{quarkprop} P.O.~Bowman, U.M.~Heller and A.G.~Williams,
	hep-lat/0203001. 

\bibitem{vBa95} P.~van Baal, 
	\npps{42}{843}{1995}.

\bibitem{Man01} J.E.~Mandula,
	\npps{106}{998}{2002}.

\bibitem{Bon02} F.D.R.~Bonnet, P.O.~Bowman, D.B.~Leinweber,
	A.G.~Williams and J.~Zhang,
	To appear in Physical Review, hep-lat/0202003.

\end{thebibliography}
\end{document}